\documentclass[sigplan,10pt]{acmart}

\usepackage[american]{babel}
\usepackage{amsmath, amssymb}
\usepackage[scaled]{beramono}
\usepackage{hyperref}
\usepackage{cleveref}
\usepackage{listings}
\usepackage{mathpartir}
\usepackage{mathtools}
\usepackage{multirow}
\usepackage{semantic}
\usepackage{standalone}
\usepackage{subcaption}
\usepackage{tabularx}
\usepackage{todonotes}

\usepackage{tikz}
\usetikzlibrary{
  calc,
  fadings,
  patterns,
  positioning,
  shapes.arrows,
  shapes.misc,
}

\usepackage{tcolorbox}
\tcbuselibrary{breakable,listings,skins}

\acmConference[MPLR'2019]{Managed Programming Languages and Runtimes}{October 20--25, 2019}{Athens, Greece}
\acmYear{2019}
\acmISBN{} % \acmISBN{978-x-xxxx-xxxx-x/YY/MM}
\acmDOI{} % \acmDOI{10.1145/nnnnnnn.nnnnnnn}
\startPage{1}

\setcopyright{none}
\definecolor{codecommentcolor}{RGB}{81,99,116}
\definecolor{codeliteralcolor}{RGB}{0,128,0}
\definecolor{codekeywordcolor1}{RGB}{180,88,172}
\definecolor{codekeywordcolor2}{RGB}{180,88,0}
\definecolor{codekeywordcolor3}{RGB}{77,77,255}

\newtcblisting{codebox}[2][]{%
  enhanced,
  breakable,
  attach boxed title to top right={xshift=-\tcboxedtitleheight/2,yshift=-\tcboxedtitleheight/2},
  boxed title style={size=small,colback=gray!5},
  coltitle=black!75,
  listing only,
  top=0mm,
  bottom=0mm,
  left=6mm,
  leftrule=1pt,
  toprule=1pt,
  rightrule=1pt,
  bottomrule=1pt,
  leftrule=1pt,
  listing options={#1},
  colframe=gray!20,
  colback=gray!5,
  #2
 }
\lstdefinelanguage{anzen} {
  keywords    = [1]{
    let, var, in, fun, new, if, then, else, return, struct, self,
    true, false,
    @mut, @cst, @own, @brw, @esc, mutating
  },
  keywords    = [2]{nothing, Int, String, Bool},
  keywords    = [3]{print},
  morecomment = [l]{//},
  morecomment = [s]{/*}{*/},
  morestring  = [b]",
}

\lstdefinelanguage{javascript} {
  keywords    = [1]{
    let, var, const, class, function, return, if, else, this
  },
  keywords    = [2]{},
  keywords    = [3]{console},
  morecomment = [l]{\#},
  morestring  = [b]",
}

\newcommand*{\tuple}[1]{\ensuremath{\langle #1 \rangle}}
\newcommand*{\card}[1]{\ensuremath{\parallel #1 \parallel}}

\newcommand*{\ident}[1]{\ensuremath{\mathit{#1}}}

\newcommand*{\dom}{\ensuremath{\mathrm{dom}}}
\newcommand*{\codom}{\ensuremath{\mathrm{codom}}}
\newcommand*{\powerset}{\ensuremath{\mathcal{P}}}

\newcommand*{\aliasop}{~\text{\texttt{\&-}~}}
\newcommand*{\moveop}{\leftarrow}
\newcommand*{\copyop}{\coloneqq}

\newcommand*{\AC}{\ensuremath{\mathcal{A}}}
\newcommand*{\Atom}{\ensuremath{\mathbb{A}}}
\newcommand*{\Capa}{\ensuremath{\mathbb{C}}}
\newcommand*{\Id}{\ensuremath{\mathbb{X}}}
\newcommand*{\Type}{\ensuremath{\mathbb{T}}}
\newcommand*{\Mem}{\ensuremath{L}}
\newcommand*{\Op}{\ensuremath{\mathbb{O}}}
\newcommand*{\Qual}{\ensuremath{\mathbb{Q}}}
\newcommand*{\Ref}{\ensuremath{R}}
\newcommand*{\Val}{\ensuremath{\mathbb{V}}}

\newcommand*{\acvar}[2]{\ensuremath{\mathtt{var}~#1~\mathtt{in}~#2}}
\newcommand*{\acret}[1]{\ensuremath{\mathtt{ret}~#1}}
\newcommand*{\acnew}[1]{\ensuremath{\mathtt{new}~#1}}
\newcommand*{\acif}[3]{\ensuremath{#1~?~#2~!~#3}}

\newcommand*{\acqual}[1]{\ensuremath{\mathtt{#1}}}
\newcommand*{\accst}{\acqual{cst}}
\newcommand*{\acmut}{\acqual{mut}}
\newcommand*{\acown}{\acqual{own}}
\newcommand*{\acbrw}{\acqual{brw}}

\newcommand*{\partialto}{\nrightarrow}

\title{Explicit and Controllable Assignment Semantics}

\author{Dimitri Racordon}
\affiliation{%
  \institution{University of Geneva}
  \department{Centre Universitaire d'Informatique}
  \country{Switzerland}
}
\email{dimitri.racordon@unige.ch}
\author{Didier Buchs}
\affiliation{%
  \institution{University of Geneva}
  \department{Centre Universitaire d'Informatique}
  \country{Switzerland}
}
\email{didier.buchs@unige.ch}

\begin{document}

\begin{abstract}
Despite the plethora of powerful software to spot bugs,
identify performance bottlenecks or simply improve the overall quality of code,
programming languages remain the first and most important tool of a developer.
Therefore, appropriate abstractions, unambiguous syntaxes and intuitive semantics are paramount
to convey intent concisely and efficiently.
The continuing growth in computing power has allowed
modern compilers and runtime system to handle once thought unrealistic features,
such as type inference and reflection,
making code simpler to write and clearer to read.
Unfortunately, relics of the underlying memory model still transpire in most programming languages,
leading to confusing assignment semantics.

This paper introduces Anzen, a programming language that aims to
make assignments easier to understand and manipulate.
The language offers three assignment operators, with unequivocal semantics,
that can reproduce all common imperative assignment strategies.
It is accompanied by a type system based on type capabilities to enforce uniqueness and immutability.
We present Anzen's features informally and formalize it by the means of a minimal calculus.
\end{abstract}

%% Generate at 'http://dl.acm.org/ccs/ccs.cfm'.
\begin{CCSXML}
<ccs2012>
<concept>
<concept_id>10011007.10011006.10011008.10011009.10011010</concept_id>
<concept_desc>Software and its engineering~Imperative languages</concept_desc>
<concept_significance>500</concept_significance>
</concept>
<concept>
<concept_id>10011007.10011006.10011008.10011024</concept_id>
<concept_desc>Software and its engineering~Language features</concept_desc>
<concept_significance>500</concept_significance>
</concept>
<concept>
<concept_id>10011007.10011006.10011039</concept_id>
<concept_desc>Software and its engineering~Formal language definitions</concept_desc>
<concept_significance>300</concept_significance>
</concept>
<concept>
<concept_id>10011007.10011006.10011041</concept_id>
<concept_desc>Software and its engineering~Compilers</concept_desc>
<concept_significance>100</concept_significance>
</concept>
</ccs2012>
\end{CCSXML}

\ccsdesc[500]{Software and its engineering~Imperative languages}
\ccsdesc[500]{Software and its engineering~Language features}
\ccsdesc[300]{Software and its engineering~Formal language definitions}
\ccsdesc[100]{Software and its engineering~Compilers}

\keywords{imperative languages, assignment, aliasing, uniqueness, immutability}

\maketitle

\section{Introduction}

Writing software programs is a difficult task.
Moreover, with the dissemination of computer-based technology in all domains, this endeavor is no longer reserved for computer experts.
Consequently, good language design is essential to the production and maintenance of software.
Thanks to the continuing growth in computing power, contemporary programming languages can sit at very high levels of abstractions, letting developers convey their intent with great expressiveness.
Unfortunately, this evolution also brings the potential for more inadvertent behaviors, as the small imperfections in these many layers of abstractions usually have deep consequences on a language's semantics.
This is particularly true for the imperative paradigm, in which the interplay between variables and memory can prove to be a prolific source of puzzling bugs.
For instance, most mainstream object-oriented programming languages have a limited support to express aliasing explicitly, blurring the distinction between object mutation and reference reassignment, and often leading to erroneous assumptions about sharing and immutability.

This paper introduces Anzen, a multi-paradigm programming language that aims to dispel the confusion that surrounds assignment semantics.
Rather than overloading a single assignment operator with different meanings, depending on its operands' types, Anzen provides three distinct operators with unequivocal semantics.
One creates aliases, another performs deep copies and the last deals with uniqueness.
The language further supports high-level concepts, such as generic types and higher-order functions,
and offers capability-based~\cite{DBLP:conf/ecoop/BoylandNR01} memory and aliasing control mechanisms that are enforced by its runtime.
We present Anzen's features informally, by the means of various examples,
and describe its operational semantics formally.
A compiler and interpreter for Anzen are distributed as an open-source software and available at \url{https://github.com/anzen-lang/anzen}.
\section{Background and Problematic}
\label{sec:background}

In the imperative paradigm, computation is described by sequences of instructions that define \emph{how} a program operates.
These instructions interact with the state of the program,
% typically stored in the memory of the machine that runs it, and
whose modifications are expressed by memory assignments.
A memory assignment is a statement of the form ``$x \coloneqq v$" that instructs the machine to write (i.e. assign) a value to a specific location in the memory.
Here, $x$ is a variable representing a memory address and $v$ is a value.
Programming languages can express vastly different data types, ranging from numbers to more abstract data structures (e.g. a binary tree).
For example, $v$ could denote a number and be stored in memory as a simple sequence of bits, but could also denote a dynamic list represented by multiple  structures chained together.
It follows that the actual semantics of ``$\coloneqq$" has to deal with data representation and properties thereof.

A language only \emph{describes} programs, whereas their executions are driven by a runtime system that is responsible to process each instruction.
This means that the details of data representation can be eluded at the language level, as long as sufficient information can be derived by the runtime system to actually operate the machine's memory.
Most systems have evolved to divide their memory into a \emph{stack}, to store local (w.r.t. a function call), small and fixed-sized values, and a \emph{heap}, to store long-lived and arbitrarily-sized data.
A variable is merely a name denoting an address in the stack,
meaning that it cannot directly refer to heap-allocated objects.
However, addresses thereof are essentially represented as numbers, and therefore do fit into the stack.
This means that heap-allocated values are always manipulated through pointers.
Modern programming languages (e.g. Python or Javascript) are able to hide these implementation constraints so that no distinction has to be made between stack and heap allocated objects.
Thus, a variable can be seen as a container for a value of some kind, without any regard for \emph{where} this value is stored.

\begin{figure}[ht]
\centering
\resizebox{\linewidth}{!}{\begin{tikzpicture}[node distance=1.25cm,>=latex]
  \tikzset{
    word/.style={anchor=south, minimum width=2.5cm, minimum height=0.75cm},
    heap/.style={circle, minimum width=0.5cm, minimum height=0.5cm, draw, thick},
  }

  % 0xe0 [a 42]
  \node[word, fill=teal!20] (foo!a) at (0, 0) {};
    \node[anchor=east] at (foo!a.west) {\small \texttt{0xe0} };
    \node[anchor=west] at (foo!a.west) {\small \texttt{a} };
    \node[anchor=east, align=right] at (foo!a.east) {\small \texttt{42} };

  % 0xdc [b 0x2a]
  \node[word, fill=teal!20] (foo!b) at (foo!a.north) {};
    \node[anchor=east] at (foo!b.west) {\small \texttt{0xdc} };
    \node[anchor=west] at (foo!b.west) {\small \texttt{b} };
    \node[anchor=east, align=right] at (foo!b.east) {\small \texttt{0x2a} };

  % 0xd8 [x 42]
  \node[word, pattern color=teal!50, pattern=north west lines] (foo!c) at (foo!b.north) {};
    \node[anchor=east] at (foo!c.west) {\small \texttt{0xd8} };
    \node[anchor=west] at (foo!c.west) {\small \texttt{x} };
    \node[anchor=east, align=right] at (foo!c.east) {\small \texttt{42} };

  % 0xd8 [y 0x2a]
  \node[word, pattern color=teal!50, pattern=north west lines] (foo!d) at (foo!c.north) {};
    \node[anchor=east] at (foo!d.west) {\small \texttt{0xd4} };
    \node[anchor=west] at (foo!d.west) {\small \texttt{y} };
    \node[anchor=east, align=right] at (foo!d.east) {\small \texttt{0x2a} };

  % Draw a frame around the stack.
  \draw[dashed, path fading=north]
    (foo!d.north west) -- ($(foo!d.north west)+(0, 0.5)$)
    (foo!d.north east) -- ($(foo!d.north east)+(0, 0.5)$);
  \draw[dashed, path fading=south]
    (foo!a.south west) -- ($(foo!a.south west)+(0, -0.5)$)
    (foo!a.south east) -- ($(foo!a.south east)+(0, -0.5)$);
  \draw[]
    (foo!a.south west) -- (foo!d.north west)
    (foo!a.south east) -- (foo!d.north east);

  % Draw some heap nodes.
  \node[heap, draw=blue!50, fill=blue!20] (h01) at ($(foo!d.north east) + (3.75,0)$) {};
  \node[heap, draw=blue!50, fill=blue!20] (h02) at ($(h01) + (-0.75,-1.25)$) {};
  \node[heap, draw=blue!50, fill=blue!20] (h03) at ($(h01) + (0.75,-1.25)$) {};
  \node[heap, draw=blue!50, fill=blue!20] (h04) at ($(h03) + (-0.75,-1.25)$) {};
  \node[heap, draw=blue!50, fill=blue!20] (h05) at ($(h03) + (0.75,-1.25)$) {};
  % \node[heap, draw=blue!50, fill=blue!20] (h05) at ($(foo!b.north east) + (3,0)$) {};
  % \node[heap, draw=blue!50, fill=blue!20] (h06) at ($(h05) + (0,1.25)$) {};
  % \node[heap, draw=blue!50, fill=blue!20] (h07) at ($(h05) + (1.25,-0.5)$) {};
  % \node[heap, draw=blue!50, fill=blue!20] (h08) at ($(h05) + (2.5,0)$) {};
  % \node[heap, draw=blue!50, fill=blue!20] (h09) at ($(h06) + (1.25,0.5)$) {};

  \draw[->, draw=gray!50] (h01) -> (h02);
  \draw[->, draw=gray!50] (h01) -> (h03);
  \draw[->, draw=gray!50] (h03) -> (h04);
  \draw[->, draw=gray!50] (h03) -> (h05);
  % \draw[->, draw=gray!50] (h05) -> (h06);
  % \draw[->, draw=gray!50] (h06) -> (h09);
  % \draw[->, draw=gray!50] (h06) -> (h08);
  % \draw[->, draw=gray!50] (h08) -> (h07);
  % \draw[->, draw=gray!50] (h05) -> (h07);

  \draw[->, thick, rounded corners] (foo!b.east) to[in=180] (h01);
  \draw[->, thick, dotted, rounded corners] (foo!d.east) to[in=160] (h01);

  \draw[dashed] ($(foo!a.south east) + (1.5,-0.5)$) -- ($(foo!d.north east) + (1.5,1.5)$) coordinate (title);
  \node[align=right, anchor=east] at ($(title) + (-0.25,-0.5)$) {\strut Stack};
  \node[align=left , anchor=west] at ($(title) + (+0.25,-0.5)$) {\strut Heap};
\end{tikzpicture}}
\caption{
Effect of bitwise copy assignment.
Two local variables \lstinline|x| and \lstinline|y| are assigned to two other local variables \lstinline|a| and \lstinline|b|, respectively.
The variable \lstinline|a| represents a number allocated on the stack.
It is copied during \lstinline|x|'s assignment, which therefore represents a different object.
On the other hand, since \lstinline|b| represents a tree allocated on the heap,
its \emph{pointer} is copied during \lstinline|y|'s assignment, resulting in an alias.
}
\label{fig:bitwise-copy}
\end{figure}
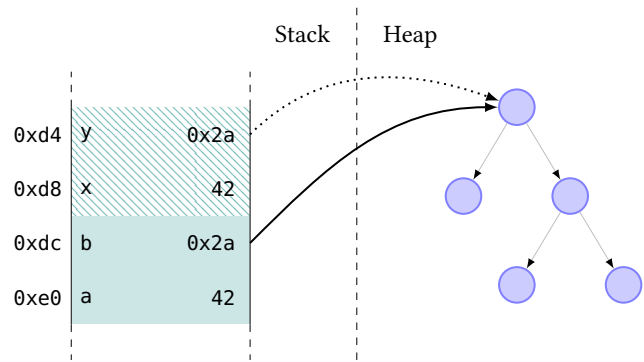

Unfortunately, most execution models do not preserve this abstraction when dealing with assignments, which are implemented as a straightforward ``bitwise copy" of the stack-allocated value a variable represents.
Consequently, values stored on the stack get fully copied, whereas values stored on the heap and referred to by a pointer get aliased,
since only the their pointers' values are copied.
\Cref{fig:bitwise-copy} illustrates both cases.
Although this strategy is efficient and usually matches the developer's intent, it may leads to confusing situations, in particular when mutation is involved.
Consider for instance the two Python snippets below.
On the left, the variable \lstinline|a|'s value is copied.
Therefore its mutation at line 3 does not impact \lstinline|x|'s value.
Conversely on the right, the variable \lstinline|a| represents a heap-allocated dynamic array, whose pointer is copied at line 3.
Consequently the mutation at line 3 also impacts the value that \lstinline|x| represents.

\begin{tcolorbox}[
colback=white, frame hidden, left=0pt, right=0pt, boxrule=0pt, frame code={}, lower separated=false, sidebyside, sidebyside gap=3mm]
\begin{codebox}[language=python]{}
a = 4
x = a
a += 2
print(a, x)
# 6 2
\end{codebox}
\tcblower
\begin{codebox}[language=python]{}
a = [4]
x = a
a += [2]
print(a, x)
# [4,2] [4,2]
\end{codebox}
\end{tcolorbox}

Another assignment strategy that has gained popularity in recent years is to ``move'' values from one variable to another, thereby avoiding any copy or alias creation.
The advantages of this so-called \emph{move semantics} are twofold.
On the performance front, it allows efficient return-by-value implementations by eluding unnecessary copies~\cite{Books/andrist/cpp}.
On the memory safety front, it allows to preserve uniqueness, thereby guaranteeing data-race safety~\cite{DBLP:conf/ifip2/Wadler90}.
For example, the Rust programming language adopts this approach as its default assignment semantics.
However, since the language can fall back to copy semantics for certain data types, it also bears the potential for similar confusing situations.

We argue that the problem does not reside in the support of multiple assignment strategies in the same programming language.
On the contrary, those allow to better express intent and offer a tighter control over memory.
Instead, the source of the confusion stems from overloading a single operator with different semantics, defined implicitly by the type of its operands, in order to cater for the abstraction of pointers.
This comes with several important drawbacks:
\begin{itemize}
  \item This steepens the learning curve for beginners, as aliasing consistently appear as a difficult aspect of programming~\cite{Tan:ICCTD2009}.
  \item This increases the risk of errors in code refactoring, as seemingly identical statements can have different side-effects.
  \item This hinders the control over pointers, as purposely creating aliases on stack-allocated values often requires to wrap them in heap-allocated objects, thereby reimplementing the abstraction.
\end{itemize}
A better approach would be to support these different strategies explicitly, in order to let the developer choose the most appropriate one and express her intent unambiguously.

\section{The Anzen Programming Language}
\label{sec:anzen}

Anzen is a typed general purpose language that aims to make assignment semantics explicit.
Although it supports multiple paradigms, Anzen leans towards object orientation and imperative programming.
We introduce the language and its features with a series of examples.

\subsection{Assignment Semantics}

Anzen's most distinguishing characteristic is that it features three different assignment operators.
These are independent of their operands' types and provide the following strategies:
\begin{itemize}
  \item An aliasing operator $\aliasop$ assigns an alias on the object on its right to the reference on its left.
    Its semantics is the closest to what is generally understood as an assignment in languages that abstract over pointers (e.g. Python or Java).
  \item A copy operator $\copyop$ assigns a \emph{deep} copy of the object's value on its right to the reference on its left.
    If the left operand was already bound to an object, the copy operator mutates its value rather than reassigning the reference to a different one.
  \item A move operator $\moveop$ moves the object on its right to the reference on its left.
    If the right operand was a reference, the move operator removes its binding, effectively leaving it unusable until it is reassigned.
\end{itemize}

\begin{figure}[ht]
  \begin{minipage}{1\linewidth}
    \centering
    \resizebox{\linewidth}{!}{\begin{tikzpicture}
      \tikzset{
        root/.style={rectangle, minimum width=0.5cm, minimum height=0.5cm, draw, thick},
        heap/.style={circle, minimum width=0.5cm, minimum height=0.5cm, draw, thick},
      }

      % Before the assignment ...

      \node[root] (a0) at (0, 0) {$a$};
      \node[root] (b0) [below=0.5cm of a0] {$b$};
      \node[heap] (m00) at ($(a0) + (2,0)$) {8};
      \node[heap] (m01) at ($(b0) + (2,0)$) {4};
      \draw[->, thick] (a0) -> (m00);
      \draw[->, thick] (b0) -> (m01);

      \draw[dashed] ($(b0) + (1,-0.5)$) -- ($(a0) + (1,1.5)$) coordinate (t0);
      \node[align=right, anchor=east] at ($(t0) + (-0.25,-0.5)$) {\strut References};
      \node[align=left , anchor=west] at ($(t0) + (+0.25,-0.5)$) {\strut Memory};

      % Assignment ...

      \node[
        single arrow, fill=black!20, minimum height=0.75cm, path fading=west,
        label={[label distance=0.25cm]above:$b ~\text{\texttt{\&-}}~ a$}
      ] (stmt0) at ($(m00)+(1.5,-0.5)$) {};

      % After the assignment ...

      \node[root] (a1) at ($(stmt0)+(1.5,0.5)$) {$a$};
      \node[root, draw=orange!50, fill=orange!20] (b1) [below=0.5cm of a1] {$b$};
      \node[heap] (m10) at ($(a1) + (2,0)$) {8};
      \node[heap] (m11) at ($(b1) + (2,0)$) {4};
      \draw[->, thick] (a1) -> (m10);
      \draw[->, draw=orange!50, thick] (b1) to[out=0, in=180] (m10);

      \draw[dashed] ($(b1) + (1,-0.5)$) -- ($(a1) + (1,1.5)$) coordinate (t1);
      \node[align=right, anchor=east] at ($(t1) + (-0.25,-0.5)$) {\strut References};
      \node[align=left , anchor=west] at ($(t1) + (+0.25,-0.5)$) {\strut Memory};
    \end{tikzpicture}}
    \subcaption{Aliasing operator}
  \end{minipage}\hfill%
  \begin{minipage}{1\linewidth}
    \centering
    \resizebox{\linewidth}{!}{\begin{tikzpicture}
      \tikzset{
        root/.style={rectangle, minimum width=0.5cm, minimum height=0.5cm, draw, thick},
        heap/.style={circle, minimum width=0.5cm, minimum height=0.5cm, draw, thick},
      }

      % Before the assignment ...

      \node[root] (a0) at (0, 0) {$a$};
      \node[root] (b0) [below=0.5cm of a0] {$b$};
      \node[heap] (m00) at ($(a0) + (2,0)$) {8};
      \node[heap] (m01) at ($(b0) + (2,0)$) {4};
      \draw[->, thick] (a0) -> (m00);
      \draw[->, thick] (b0) -> (m01);

      \draw[dashed] ($(b0) + (1,-0.5)$) -- ($(a0) + (1,1.5)$) coordinate (t0);
      \node[align=right, anchor=east] at ($(t0) + (-0.25,-0.5)$) {\strut References};
      \node[align=left , anchor=west] at ($(t0) + (+0.25,-0.5)$) {\strut Memory};

      % Assignment ...

      \node[
        single arrow, fill=black!20, minimum height=0.75cm, path fading=west,
        label={[label distance=0.25cm]above:$b \coloneqq a$}
      ] (stmt0) at ($(m00)+(2,-0.5)$) {};

      % After the assignment ...

      \node[root] (a1) at ($(stmt0)+(2,0.5)$) {$a$};
      \node[root] (b1) [below=0.5cm of a1] {$b$};
      \node[heap] (m10) at ($(a1) + (2,0)$) {8};
      \node[heap, draw=orange!50, fill=orange!20] (m11) at ($(b1) + (2,0)$) {8};
      \draw[->, thick] (a1) -> (m10);
      \draw[->, thick] (b1) -> (m11);

      \draw[dashed] ($(b1) + (1,-0.5)$) -- ($(a1) + (1,1.5)$) coordinate (t1);
      \node[align=right, anchor=east] at ($(t1) + (-0.25,-0.5)$) {\strut References};
      \node[align=left , anchor=west] at ($(t1) + (+0.25,-0.5)$) {\strut Memory};
    \end{tikzpicture}}
    \subcaption{Copy operator}
  \end{minipage}\hfill%
  \begin{minipage}{1\linewidth}
    \centering
    \resizebox{\linewidth}{!}{\begin{tikzpicture}
      \tikzset{
        root/.style={rectangle, minimum width=0.5cm, minimum height=0.5cm, draw, thick},
        heap/.style={circle, minimum width=0.75cm, minimum height=0.5cm, draw, thick},
        cross/.style={cross out, draw=red, minimum size=0.25cm, inner sep=0pt, outer sep=0pt},
      }

      % Before the assignment ...

      \node[root] (a0) at (0, 0) {$a$};
      \node[root] (b0) [below=0.5cm of a0] {$b$};
      \node[heap] (m00) at ($(a0) + (2,0)$) {8};
      \node[heap] (m01) at ($(b0) + (2,0)$) {4};
      \draw[->, thick] (a0) -> (m00);
      \draw[->, thick] (b0) -> (m01);

      \draw[dashed] ($(b0) + (1,-0.5)$) -- ($(a0) + (1,1.5)$) coordinate (t0);
      \node[align=right, anchor=east] at ($(t0) + (-0.25,-0.5)$) {\strut References};
      \node[align=left , anchor=west] at ($(t0) + (+0.25,-0.5)$) {\strut Memory};

      % Assignment ...

      \node[
        single arrow, fill=black!20, minimum height=0.75cm, path fading=west,
        label={[label distance=0.25cm]above:$b \leftarrow a$}
      ] (stmt0) at ($(m00)+(2,-0.5)$) {};

      % After the assignment ...

      \node[root, draw=orange!50, fill=orange!20] (a1) at ($(stmt0)+(2,0.5)$) {$a$};
      \node[root] (b1) [below=0.5cm of a1] {$b$};
      \node[heap, dotted] (m10) at ($(a1) + (2,0)$) {};
      \node[heap, draw=orange!50, fill=orange!20] (m11) at ($(b1) + (2,0)$) {8};
      \draw[->, draw=orange!50, thick] (a1) -- ($(a1) + (0.75,0)$) node[cross] {};
      \draw[->, thick] (b1) -> (m11);

      \draw[dashed] ($(b1) + (1,-0.5)$) -- ($(a1) + (1,1.5)$) coordinate (t1);
      \node[align=right, anchor=east] at ($(t1) + (-0.25,-0.5)$) {\strut References};
      \node[align=left , anchor=west] at ($(t1) + (+0.25,-0.5)$) {\strut Memory};
    \end{tikzpicture}}
    \subcaption{Move operator}
  \end{minipage}
  \caption{
    Effect of assignment operators.
    Each illustration depicts the situations before and after a particular assignment,
    starting from a state where two variables $a$ and $b$ are bound to unrelated memory locations,
    holding the values 8 and 4 respectively.
    Changes are highlighted in color.}
  \label{fig:assignment-operators}
\end{figure}
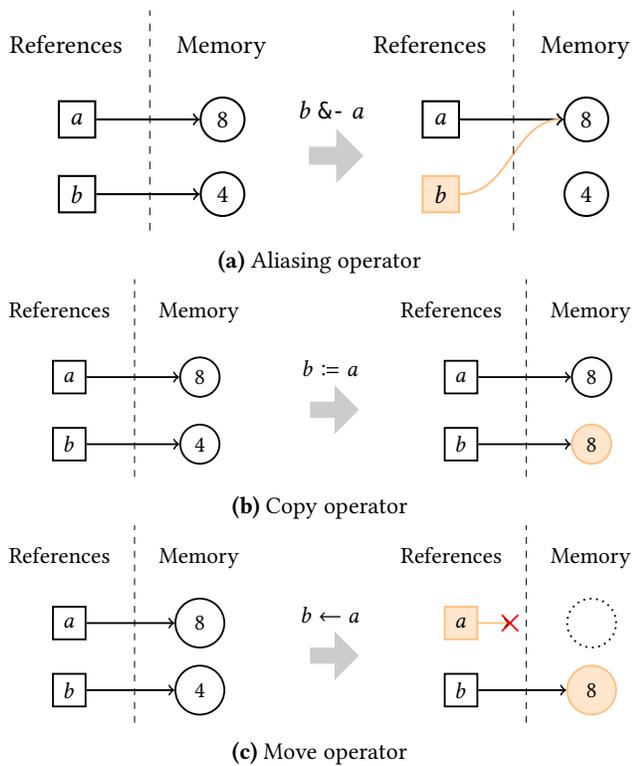

\Cref{fig:assignment-operators} illustrates the three assignment semantics.
Notice that we use the term ``memory'' rather than ``stack'' or ``heap''.
The reason is that we aim at making a clear distinction between the semantics of Anzen's operators and the actual memory model that is used to implement them.
Whether a reference is a pointer to heap-allocated memory or a simple value allocated on the stack should be irrelevant for the developer.
That way she may focus solely on the semantics of her program, at an abstraction level that does not bother about the specifics of compilation, optimization and/or interpretation.

\subsection{Parameter Passing}

There is a tight connection between reference assignments and the passing of parameters in a function call~\cite{DBLP:conf/popl/OderskyRH93}, as one can see a parameter as a mere variable assigned to a corresponding argument.
Consequently, just as most runtime systems implement assignments as a straightforward ``bitwise copy'', they often do the same for function parameters, which are passed as a bitwise copy the argument's value.
This makes sense from an implementation perspective, as it preserves the symmetry with assignment semantics.
Unfortunately, it may also cause similar confusing situations as those we described in \Cref{sec:background}.
% , with an extra layer of complexity introduced by the transfer of control flow.
For instance, although it is technically correct to state that Java adopts a pass-by-value semantics, as mentioned in the language manual~\cite[Chapter~4]{Gosling:2019:JLS}, the observable behavior is a pass-by-alias (a.k.a. pass-by-reference) for all heap-allocated objects, as mutating such parameters do have side effects on the call site.
Anzen fends off this issue by reusing its three assignment operators to specify the parameter passing strategy explicitly.
The use of an aliasing operator corresponds to a pass-by-alias, while the use of a copy operator corresponds to a pass-by-value policy. Unsurprisingly, the move operator also treats objects as linear resources, and therefore mimics Rust's move semantics.
An example follows:

\begin{codebox}[language=anzen]{title={Parameter passing}}
fun f(list: @mut List<Int>) {
  list.append(new_element <- 5)
  print(line <- list)
}

var a: @mut <- List<Int>()
a.append(new_element <- 4)

// `a` is passed by value, therefore the
// mutation of the `list` parameter in
// `f`'s body have no side effects.
f(list := a)
// Prints "[4, 5]"

print(line <- a)
// Prints "[4]"
\end{codebox}

 Reusing the assignment operators makes the relationship between assignment and parameter passing explicit, which we believe to be beneficial for educational purposes.
 Even the induced syntactic overhead may in fact be seen as an improvement.
 Indeed, whereas it is traditional to identify parameters by their position, one can leverage named parameters to improve the legibility of the code. For instance, the expression \lstinline|iter(from <- 0, by_steps_of <- 2)| is arguably more self-explanatory than \lstinline|iter(0, 2)|.
 This peculiarity is often referred to as ``named parameters'' in other languages (e.g. in Python).

 For the sake of code legibility, arithmetic operators support an infix notation, that assumes a pass-by-value strategy for each operand.
 For example, the expression \lstinline|2 + 3| is a syntactic sugar for \lstinline|Int.+(lhs := 2, rhs := 3)|.

 \subsection{Data Types}

 On the top of a handful of primitive types to describe numbers and boolean values, Anzen also supports custom type definitions in the form of structures.
 A structure is an aggregate of named fields, which represent references to other values.
 Borrowing from object-oriented programming languages, methods, constructors and destructors can be declared along with a structure to compartmentalize behavior.
 Inside these functions, a reserved \lstinline|self| reference allows access to the instance's internals.
 The language does not however support inheritance and polymorphism.

% \begin{codebox}[language=anzen]{title={Structures}}
% struct Point2D {
%   var x: Float
%   var y: Float

%   fun distance(other: Point2D) -> Float {
%     var dx <- self.x - other.x
%     var dy <- self.y - other.y
%     return <- sqrt(x <- dx * dx + dy * dy)
%   }
% }
% \end{codebox}

Anzen is a higher-order programming language.
Any identifier that is neither a parameter nor a variable declared within the function's body is captured by closure.
Captured identifiers are in fact aliases on the corresponding variables in the function's declaration context.
It follows that mutating their values will also modify the value within the closure.

% \begin{codebox}[language=anzen]{title={Higher-order functions}}
% var counter: @mut <- 0
% fun next_number() -> Int {
%   counter <- counter + 1
%   return := counter
% }

% print(line <- next_number())
% // Prints "1"
% print(line := counter)
% // Prints "1"
% counter := 10
% print(line <- next_number())
% // Prints "11"
% print(line := counter)
% // Prints "11"
% \end{codebox}

\section{Aliasing Control Mechanisms}

Anzen's type system offers some aliasing control mechanisms to enforce uniqueness and immutability at runtime.
This section introduces them informally.

\subsection{Ownership and Uniqueness}

Uniqueness~\cite{DBLP:conf/oopsla/GordonPPBD12} is a property of a reference that holds if there are no other references (i.e. aliases) on the object to which it is bound.
In order to avoid dangling references, the right operand of a move operator has to be unique,
otherwise aliases would refer to a moved object.
Therefore Anzen's type system has to keep track of reference uniqueness.
This is achieved by the means of a typestate analysis~\cite{DBLP:journals/tse/StromY86}.
% This is achieved by the means of type capabilites~\cite{DBLP:conf/ecoop/BoylandNR01}.

We first define a notion of object ownership.
A reference is said to bear ownership on the object to which it is bound if it has been initialized using either the copy or the move operator.
Notice that this means an owning reference has necessarily been unique at least once during the program's execution.
Then, we define the following five reference states:
the \emph{unallocated} state denotes references that are not bound to any object,
the \emph{unique} state denotes owning references to unaliased objects,
the \emph{shared} state denotes owning references to aliased objects,
the \emph{borrowed} state denotes non-owning references, and
the \emph{moved} state denotes references that have been moved.
% A reference always starts in the unallocated state.
% If assigned by copy or move, it receives ownership and becomes unique.
% If assigned by alias, it becomes borrowed.
% A unique reference becomes shared after appearing at the right side of an aliasing operator.
% It may go back to the unique state if all aliases are either reassigned or go out of scope.
% Finally, a unique reference goes to the moved state after appearing at the right side of a move operator.
\Cref{tab:reference-typestate} summarizes how these transitions can take place.
Each cell represents the state in which a reference goes if placed on the left or right side of the operator indicated by the column.
The position of the bullet denotes the side of the operator on which the reference should be.
For instance, $\bullet\copyop$ denotes the left operand of a copy assignment.
An example follows:

\begin{codebox}[language=anzen]{title={State transitions}}
var a, b, c: Int
// `a`, `b` and `c` start unallocated
a := 10
b <- 20
// `a` and `b` get unique
c &- a
// `a` gets shared, `c` gets borrowed
c &- b
// `b` gets shared, `a` gets unique
b <- a
// `a` gets moved
\end{codebox}

\begin{table*}
  \centering%
  \begin{tabular}{ r|c|c|c||c|c|c| }
    \multicolumn{1}{c|}{\multirow{2}{*}{$\nearrow$}} & \multicolumn{3}{c||}{as left operand} & \multicolumn{3}{c|}{as right operand} \\ \cline{2-7}
    & $\bullet\aliasop$ & $\bullet\copyop$ & $\bullet\moveop$ & $\aliasop\bullet$ & $\copyop\bullet$ & $\moveop\bullet$ \\ \hline
    unalloc. & borrowed & unique & unique & \textcolor{red}{$\times$} & \textcolor{red}{$\times$} & \textcolor{red}{$\times$} \\ \hline
    unique   & borrowed & unique & unique & shared & unique & moved \\ \hline
    shared   & \textcolor{red}{$\times$} & shared & shared & shared & shared & \textcolor{red}{$\times$} \\ \hline
    borrowed & borrowed & borrowed & borrowed & borrowed & borrowed & \textcolor{red}{$\times$} \\ \hline
    moved    & borrowed & unique & unique & \textcolor{red}{$\times$} & \textcolor{red}{$\times$} & \textcolor{red}{$\times$} \\ \hline
  \end{tabular}
  \caption{Reference state transitions}
  \label{tab:reference-typestate}
\end{table*}

Uniqueness is treated as a fractionable capability, reminiscent to Boyland's proposal~\cite{DBLP:conf/sas/Boyland03}.
The intuition is that it is initially obtained in full at allocation, but gets fractioned for each alias.
The loss of the uniqueness capability is only temporary, as an owner can later gather all fragments back to rebuild it.
Put another way, each aliasing assignment creates a new fragment of uniqueness, while one is sent back to its owner every time an alias goes out of scope or is reassigned.
Aliases themselves are allowed to further fraction their fragment to allow additional aliases.
However this does not create a hierarchy, and a fragment is always directly sent back to its owner.
This mechanism effectively implements borrowing and reborrowing~\cite{DBLP:conf/popl/NadenBAB12}.

As the tracking of uniqueness fragments is performed at runtime, our model does not require an elaborate pointer analysis and does not have to deal with statically nondeterministic situations (e.g. conditional expressions).
Instead the mechanism can be implemented by the means of a simple table associating references to capabilities at runtime.
Incidentally, this explains why shared references cannot be reassigned by alias.
The restriction guarantees that borrowed references can properly return their uniqueness fragment, without any additional bookkeeping.

\subsection{References and Mutability}

The term ``immutability'' can be a placeholder for vastly different concepts, depending on the programming language.
We briefly describe them to establish the vocabulary we use in the remainder of this paper.
A more comprehensive discussion about immutability is given in~\cite{DBLP:series/lncs/PotaninOZE13}.

\begin{description}
  \item[Non-reassignability]
    is a property of references, indicating that they cannot be reassigned to another object after having being assigned once.
    This is the most common form of immutability found in programming languages that abstract over heap memory management (e.g. Javascript's \lstinline[language=javascript]|const| or Java's \lstinline[language=java]|final|).
  \item[Reference immutability]
    is a property of references, indicating that the object to which they refer cannot be modified \emph{through} them.
    In other words, an immutable reference is an alias through which the object's mutation is not allowed.
  \item[Object immutability]
    is a property of objects, indicating that they cannot be modified, by any reference.
    The property can be applied on the bits that constitute the object only, or also transitively applied to the other objects to which it refers.
    The former strategy, called \emph{shallow immutability}, only prevents fields from being reassigned, whereas the latter, called \emph{deep immutability}, freezes the entire object's representation.
\end{description}

Anzen supports non-reassignable references and deep object immutability.
Non-reassignable references are declared with the keyword \lstinline|let|, whereas reassignable references are declared with the keyword \lstinline|var|.
Object immutability on the other hand is specified with type qualifiers.
\lstinline|@cst| describes immutable values whereas \lstinline|@mut| describes a mutating one.
We say that a reference to an immutable (resp. mutable) object is \emph{non-mutating} (resp. \emph{mutating}).
\Cref{tab:variable-declaration} illustrates all possible combinations of reassignability and object immutability.
Anzen chooses to always prefer immutability over mutability.
Accordingly, omitting the mutability qualifier is interpreted as a \lstinline|@cst| by default, and parameters are always assumed non-reassignable.

\begin{table}[ht]
  \centering%
  \begin{tabular}{ r|c|c| }
    & \multicolumn{1}{c|}{Reassignable} & \multicolumn{1}{c|}{Non-reassignable} \\ \hline
    Mutating     & \lstinline[language=anzen]|var a: @mut| & \lstinline|let b: @mut| \\ \hline
    Non-mutating & \lstinline[language=anzen]|var c: @cst| & \lstinline|let d: @cst| \\ \hline
  \end{tabular}
  \caption{References and mutability}
  \label{tab:variable-declaration}
  \vspace{-1em}
\end{table}

Reassignability is checked statically, meaning that reassigning a reference declared with \lstinline|let| or a function parameter is rejected during compilation.
On the contrary, object mutability is checked dynamically, meaning that modifying an immutable object triggers an error during execution.

\begin{codebox}[language=anzen]{title={Immutability}}
let a: @cst <- 42
a := 10
// At runtime: `a` is not mutating

let b: @cst <- 1337
a &- b
// At compilation: `a` is not reassignable
\end{codebox}

The type system guarantees that mutating references cannot coexist with non-mutating references on the same object.
However, unlike in most systems, there is no restriction on the number of mutating references.
The reasons are twofold.
Firstly, this allows mutable self-referential data structures to be expressed naturally, whereas such an exercise can prove challenging in more constrained type systems~\cite{DBLP:conf/sosp/LevyACCDGLP15}.
Secondly, Anzen is currently single-threaded, and therefore not susceptible to data races in the presence of multiple mutating references.
Single-threading does not discard concurrency, which is planned to be implemented by the means of cooperative multitasking~\cite{DBLP:journals/toplas/MouraI09} in a future language extension.
Cooperative multitasking differs the more widespread preemptive counterpart in that control is transferred voluntarily between concurrent processes, offering synchronization for free and therefore avoiding data race issues.

Mutual exclusion of mutating and non-mutating references is achieved by associating them with additional capabilities.
Two are defined: the \emph{read-write} and the \emph{read-only} capabilities.
At initialization non-mutating references (i.e. those annotated with \lstinline|@cst|) receive the read-only capability,
whereas mutating references (i.e. those annotated with \lstinline|@mut|) receive both.
Mutability capabilities are borrowed with the aliasing operator (i.e. \lstinline|&-|).
In order to avoid inconsistent situations, aliasing is only allowed in two situations:
\begin{itemize}
  \item
    If the left operand's type is qualified by \lstinline|@mut|, the right operand should be mutating and not aliased by non-mutating references.
  \item
    If the left operand's type is qualified by \lstinline|@cst|, the right operand should be non-mutating and/or unique.
\end{itemize}
The latter case allows unique mutating references to be shared non-mutating temporarily (e.g. for the duration of a function call).
Recall that mutating references receive both mutability capabilities upon allocation.
Hence they can lend the read-only capability to a non-mutating alias, and refrain from using their read-write capability for the duration of the loan.
An alias only holds the capability it borrows, thus preventing it to be reborrowed with an incompatible mutability.
The reader will notice that this model shares a few similarities with the notion of uniqueness that we have discussed in the previous section.
Unique references fraction either of their mutability properties when aliased,
and these fragments are sent back their owner when borrowed references are either reassigned or go out of scope.

\lstinline|self| is always a non-reassignable reference, but whether or not it designates an immutable object depends on the method's declaration.
By default, methods cannot mutate their internals via \lstinline|self|, but their definition can be prefixed with \lstinline|mutating| to turn \lstinline|self| into a mutating reference.
Object immutability is applied transitively.
In oder words, mutating a field of an immutable structure instance is an illegal operation, and provokes an error at runtime.
Consequently, calling a mutating method on a reference that is not typed with \lstinline|@mut| is forbidden, as demonstrated below:

\begin{codebox}[language=anzen]{title={Deep immutability}}
struct Point {
  // ...
  mutating fun move_x(dx: Float) {
    self.x <- self.x + dx
  }
}

let pt: @cst <- Point(x <- 1.0, y <- 1.0)
// `pt` is declared non-mutating
pt.move_x(dx <- 3.14)
// At runtime: `pt` is not mutating
\end{codebox}

\subsection{Restrictions for Code Legibility}

The mechanisms that we have described so far are sufficient to keep track of uniqueness and immutability at runtime.
However, performing the same task can prove very difficult for a developer, since our model offers a very limited syntactic support to determine where uniqueness and immutability properties are obtained, lost and retrieved.
To remedy this situation, we add a few restrictions and annotations to help the developer infer the references' typestate transitions.

First, functions and method signatures must be annotated to specify arguments and return values' expected ownership.
The \lstinline|@own| qualifier denotes and is assumed by default, while the \lstinline|@brw| qualifier denotes borrowed references.
% Although these annotations are not strictly necessary, they can help the developer reason about the kind of references she can expect at function boundaries.
The runtime system raises an error if they are not honored.
For instance, the result of a call to a function whose codomain is annotated with \lstinline|@own| should be safely consumed by move.
Consequently, returning a value by alias in such a function is illegal.

\begin{codebox}[language=anzen]{title={Ownership annotations}}
fun f(x: @own Int) -> @own Int {
  return &- x
  // At runtime: return value must be
  // passed with ownership
}
\end{codebox}

We also prevent functions from forming aliases on their arguments, unless those are annotated with \lstinline|@esc|.
This mechanism is used to prevent a function from implicitly creating a non-mutating reference on an argument that would live longer than the function call, as this would hinder the ability of a developer to determine when uniqueness and mutability can be expected to be recovered.
The constraint applies on values returned by alias and on aliases formed on captured identifiers.
Note that escaping parameters are necessarily expected to be passed by alias.

\begin{codebox}[language=anzen]{title={Escaping annotations}}
var i <- 0
fun f(x: @brw Int, y: @esc Int) {
  if x > y {
    i &- x
    // At runtime: `x` cannot escape
  } else {
    i &- y
    // Perfectly legal
  }
}
\end{codebox}

Another restriction prevents methods from returning non-mutating references on mutating fields.
As non-mutating references can assume the object to which they refer remains immutable for its entire lifetime, any mutation of the aliased field has to be forbidden.
However, there is no way to ``see'' where the non-mutating borrowing took place or where it ends from within a method.
Consequently, it also follows that a method cannot pass an alias to a mutating field to a parameter annotated with \lstinline|@esc|.
The restriction is enforced on \lstinline|self| as well, for the same reason.

\section{Example}

We now present an example of Anzen's use to showcase the advantages of its unambiguous assignment semantics and of its aliasing control mechanisms.
We use the so-called ``Signing Flaw'' security breach of JDK 1.1 as a point of reference.
In Java, all class instances (i.e. instances of \lstinline|java.lang.Class|) hold an array of signers, which acts as digitally signed identities that Java's security architecture uses to determine the class' access rights at runtime.
The value of this array can be obtained by calling \lstinline|java.lang.Class.getSigners|.
Before the breach was discovered, the method used to simply return an alias on its array of signers.
Because the returned array was an alias, an untrusted applet could freely mutate it and add trusted signers to its class, therefore fooling the security system into giving it access to restricted code signed by a trusted authority.
The following listing is an excerpt of the flawed implementation:

\begin{codebox}[language=java]{title={Flawed implementation}}
public class Class {
  public Identity[] getSigners() {
    return this.signers;
  }
  private Identity[] signers;
}
\end{codebox}

Note that none of the standard protection mechanisms of Java can help solving this issue~\cite{DBLP:journals/spe/VitekB01}.
Although the property is declared private, its reference is exposed through a method, which breaks the encapsulation principle.
Using Java's non-reassignability (i.e. using Java's \lstinline[language=java]|final| keyword) would not be of any more assistance.
While it would prevent the property from being reassigned, it would not impose any restriction on the array object itself which, consequently, could still be freely mutated.

There are two ways to prevent this situation in Anzen\footnote{%
Admittedly, Anzen does not currently support access modifiers (e.g. private), therefore the language is unable to protect the object's internal representation and forbid a direct access to the field.
We only describe protection mechanisms on the method's return value.}.
The first is to return a deep copy of signers's value by the means of the mutation operator, therefore removing any potential alias to the internal representation.

\begin{codebox}[language=anzen]{title={Using copies}}
struct Class {
  fun get_signers()
    -> @mut List<Identity>
  {
    return := self.signers
  }
  let signers: @mut List<Identity>
}
\end{codebox}

The drawback of this first approach is that it involves a deep copy, potentially expensive in terms of time and/or space.
Although optimization techniques such as copy-on-write~\cite{DBLP:conf/popl/TozawaTOM09} can alleviate this issue, a better angle would be to use a non-mutating alias.
However, as returning non-mutating aliases on mutating fields is prohibited, one cannot simply return \lstinline|signers| by alias (i.e. replacing \lstinline|:=| with \lstinline|&-| at line 5 in the above example).
Instead, one solution is to rely on Anzen's support for higher-order functions and generic types to inverse control, and accept a function that manipulates an immutable reference to the field.
That way, the non-mutating alias can be guaranteed not to escape the method's scope.

\begin{codebox}[language=anzen]{title={Using non-mutating references}}
struct Class {
  fun do_with_signers<R>(
    f: (s: @brw List<Identity>) -> @own R)
    -> @own R
  {
    return <- f(s &- self.signers)
  }
  let signers: @mut List<Identity>
}

let class = Class(signers <- List())
let count <- class.do_with_signers(
  f <- fun(s) { return <- s.count })
print(line <- count)
// Prints "0"
\end{codebox}

In the above example, the method \lstinline|do_with_signers| accepts a parameter \lstinline|f| that is a higher-order function taking a non-mutating reference on a list of identities.
It is called within \lstinline|do_with_signers|, which passes it a non-mutating reference on the \lstinline|signers| field.
Since the argument \lstinline|s| is not escaping, the non-mutating loan of the field is guaranteed to end once the function returns.
Notice that the method has a generic parameter \lstinline|R|, which types its codomain and that of the higher-order function.
This allows the former to be agnostic of the latter's return value.
The \lstinline|@own| on \lstinline|f|'s codomain guarantees that the move operator at line 6 is safe.
The method is called with an anonymous function at line 12 that reads and returns the number of signers in the list.
\section{The \texorpdfstring{$\mathcal{A}$}{A}-calculus}

We model Anzen with a minimal calculus based on a variant of the $\lambda$-calculus with assignments~\cite{DBLP:conf/popl/OderskyRH93}, which we extend with terms specific to the assignment operators we presented in \Cref{sec:anzen}.
We call such model the $\mathcal{A}$-calculus (pronounced \emph{assignment calculus}).
Programs are expressed by a single term $t \in \AC$, where $\AC$ is the set of $\mathcal{A}$-calculus' terms.

For spatial reasons, we focus on the mechanisms that are necessary to track uniqueness and object immutability.
Hence we fo not formalize reference reassignability, nor the restrictions for code legibility.

\paragraph*{Notations}
Before we delve into the syntax and semantics of the $\mathcal{A}$-calculus, we first describe the notations we will use throughout this section.
Let $\Sigma$ be a set of letters, $\Sigma^{*}$ denotes the set of all words over $\Sigma$.
Let $\overline{w} \in \Sigma^{*}$ be a word, $\card{\overline{w}}_s$ denotes the the number of occurrences of a letter $s$ in $\overline{w}$ (e.g. $\card{aab}_a = 2$).
We write $\Sigma^{\#} \subseteq \Sigma^{*}$ the set of words over $\Sigma$ without repetition.
That is $\overline{w} \in \Sigma^{\#} \Leftrightarrow \forall s \in \Sigma, \card{\overline{w}}_s \le 1$.
We write $\overline{w}_1 + \overline{w}_2$ the concatenation of two words and we write $\epsilon$ for the empty word.
Let $f$ be a function, $\dom(f)$ and $\codom(f)$ denote its domain and codomain, respectively.
The notation $f: A \to B$ denotes a total function from a domain $A$ to a codomain $B$.
The notation $g: A \partialto B$ denotes a \emph{partial} function, characterized by $\dom(f) \subseteq A$.
We overload the empty set symbol $\varnothing$ to denote partial functions with empty domains.
Let $A = \{ a_1 \dots a_n \}$ be a set of pre-images and $\{ b_1 \dots b_n \} \subseteq B$ be a set of images.
The notation $\tuple{a_1 \mapsto b_1 \dots a_n \mapsto b_n}$ denotes a function $f: A \to B$ such that $\forall_{i=1}^{n}, f(a_i) = b_i$.
Similarly, let the notation $\{ a_i \mapsto b_i \dots a_j \mapsto b_j \}$ denotes a partial function $f: A \partialto B$ such that $1 \le i \le j \le n \land \{ a_i \dots a_j \} \subseteq A$.
We borrow the so-called ``dot-notation'' from object-oriented programming language and sometimes write $f.a$ short for $f(a)$.
Let $f$ be a (total or partial) function from $A$ to $B$, we write $f[a \mapsto b]$ the \emph{update} of $f$, which is a function that maps $a$ to $b$ but $x$ to $f.x$ for any other $x \in \dom(f)$.
More formally:
\[
\forall x \in \dom(f), f[a \mapsto b] = \begin{cases}
  b &\text{if } x = a \\
  f.x &\text{otherwise}
\end{cases}
\]
We sometimes write updates by intension.
For instance, the notation $f[a \mapsto b \mid p(b)]$ denotes an update for each pre-image $a$ satisfying a predicate $p$.
Let $f$ be a (total or partial) function whose codomain is another (total or partial) function.
We write $f[a.b \mapsto c]$ short for $f[a \mapsto f.a[b \mapsto c]]$.

\subsection{Types}

The type of a term describes two kind of information.
One relates to the data type the term represent (e.g. numbers).
We call this information \emph{flow-insensitive}, since it does not depend on the evaluation context at runtime.
The other relates to typestate.
It is of course linked to the evaluation context, and is therefore termed \emph{flow-sensitive}.
Formalizing well-typedness with respect to flow-insenstive typing is beyond the scope of this paper.
For a language as simple as the $\mathcal{A}$-calculus, this fits the standard Hindley-Milner system~\cite{DBLP:journals/jcss/Milner78}.
Nonetheless, type signatures are required to type check uniqueness and immutability constraints.

\begin{definition}[Type qualifiers]
  The set of type qualifiers is given by $\Qual = \Qual_R \cup \Qual_M$, where:
  \begin{itemize}
    \item $\Qual_R = \{ \acbrw, \acown \}$ denotes reference qualifiers, and
    \item $\Qual_M = \{ \accst, \acmut \}$ denotes mutability qualifiers.
  \end{itemize}
\end{definition}

\begin{definition}[Well formed qualifier sets]
  A set of qualifiers $Q \in \powerset(\Qual)$ is \emph{well-formed} if it contains exactly one reference qualifier and exactly one mutability qualifier.
\end{definition}

\begin{definition}[Types]
  Assume a set of primitive values (e.g. numbers and booleans), written $\Atom$.
  Assume a countable set of  identifiers, written $\Id$.
  The set of types $\Type$ is defined recursively as the minimal set such that:
  \begin{itemize}
    \item
      $\tuple{q \mapsto Q, t \mapsto \Atom} \in \Type$ is an atomic type, where $Q \in \powerset(\Qual)$ is a well-formed set of qualifiers.
    \item
      $\tuple{q \mapsto Q, t \mapsto \tuple{dom \to X \to \Type, codom \to \Type}} \in \Type$ is a function type, where $Q \in \powerset(\Qual)$ is a well-formed set of qualifiers and $X \subseteq \Id$ is a set of parameter names.
    \item
      $\tuple{q \mapsto Q, t \mapsto X \to \Type} \in \Type$ is a structure type, where $Q \in \powerset(\Qual)$ is a well-formed set of qualifiers and $X \subseteq \Id$ is a set of field names.
    \item
      $\mu\alpha.\tau$ is a recursive type, where $\tau \in \Type$.
  \end{itemize}
\end{definition}

\subsection{Abstract Syntax}

\begin{definition}[Abstract syntax]
  Let $\Atom$ be the set of atomic literals and $\Id$ be the set of identifiers.
  Let $\Op = \{ \aliasop, \,\moveop\,, \,\copyop\, \}$ denote the set of assignment operators.
  The set of terms $\AC$ is defined recursively as the minimal set such that:
  \[
    \begin{array}{l l}
    a &\in \AC, \text{where } a \in \Atom \\
    x &\in \AC, \text{where } x \in \Id \\
    t.x &\in \AC, \text{where } t \in \AC \land x \in \Id \\
    \acnew{\tau} &\in \AC, \text{where } \tau \in \Type \\
    \acvar{x:\tau}{t} &\in \AC, \text{where } x \in \Id \land \tau \in \Type \land t \in \AC \\
    \lambda:\tau~\overline{x} \bullet t &\in \AC, \text{where } \tau \in \Type \land \overline{x} \in \Id^{\#} \land t \in \AC \\
    x \lhd t &\in \AC, \text{where } \lhd \in \Op \land x \in \Id \land t \in \AC \\
    t_1.x \lhd t_2 &\in \AC, \text{where } \lhd \in \Op \land x \in \Id \land t_1, t_2 \in \AC \\
    t(\overline{u}) &\in \AC, \text{where } t \in \AC \land \overline{u} \in (\Id \times \Op \times \AC)^{\#} \\
    \acret{\lhd t} &\in \AC, \text{where } \lhd \in \Op \land t \in \AC \\
    \acif{t_1}{t_2}{t_3} &\in \AC, \text{where } t_1, t_2, t_3 \in \AC \\
    t_1~;~t_2 &\in \AC, \text{where } t_1, t_2 \in \AC
    \end{array}
  \]
%   \[
%     \begin{array}{r c l}
%     a \in \Atom \implies& a &\in \AC \\
%     x \in \Id \implies& x &\in \AC \\
%     t \in \AC \land x \in \Id \implies& t.x &\in \AC \\
%     \overline{x} \in \Id^{\#} \implies& \acnew{\overline{x}} &\in \AC \\
%     x \in \Id \land t \in \AC \implies& \acvar{x}{t} &\in \AC \\
%     \overline{x} \in \Id^{\#} \land t \in \AC \implies& \lambda \overline{x} \bullet t &\in \AC \\
%     \lhd\,\in \Op \land x \in \Id \land t \in \AC \implies& x \lhd t &\in \AC \\
%     t \in \AC \land \overline{u} \in (\Id \times \Op \times \AC)^{\#} \implies& t(\overline{u}) &\in \AC \\
%     \lhd\,\in \Op \land t \in \AC \implies&\acret{\lhd t} &\in \AC \\
%     t_1, t_2, t_3 \in \AC \implies& \acif{t_1}{t_2}{t_3} &\in \AC \\
%     t_1, t_2 \in \AC \implies& t_1~;~t_2 &\in \AC
%     \end{array}
%   \]
\end{definition}

% Although parameter and argument lists are represented by words, we sometimes use commas to better distinguish between each individual subterm.
% For instance, we  write $\lambda:\tau x, y \bullet t$ rather than $\lambda:\tau xy \bullet t$.
All functions are anonymous but can be assigned to variables.
For instance we write $\acvar{f}{f \moveop \lambda:\tau~\overline{x} \triangleright t}$ to bind the variable $f$ to a function.
Consequently, recursion must be expressed by the means of the Y combinator.
We do not deal with name shadowing.
Instead we assume that all identifiers are declared only once in a given lexical scope.

Scopes are delimited implicitly.
% We sometimes use indentation to better visualize logical containment in the abstract syntax, although this is a purely stylistic choice that does not otherwise impact neither the syntax nor the semantics of a term.
Variables cannot be declared and bound to a value at the once.
Instead, variable declarations merely introduce a name into a new scope, in which initialization must be carried out.
We use parenthesis liberally to delimit scopes and clarify precedence.
Otherwise, sequences of terms are always assumed to be declared in the innermost scope, for instance both assignments are enclosed in the function's body in $\lambda:\tau~x \bullet x \moveop 2; x \moveop 3$.
Function application is left associative, for instance $f(x \moveop 1)(y \moveop 2)(z \moveop 3) = (f(x \moveop 1)(y \moveop 2))(z \moveop 3)$.
In contrast, the sequence operator $;$ is right associative, that is $t_1~;~t_2~;~t_3 = t_1~;~(t_2~;~t_3)$.

We bring the reader's attention on the fact that return expressions are made explicit.
This is a departure from most variants of $\lambda$-calculi (and functional languages in general), in which return values typically correspond the evaluation of a function's body.
We cannot take the same direction, because our syntax needs to make the return policy explicit, by the means of our assignment operators.
Furthermore, although a sequences of return statements are not syntactically ill-formed, the actual return value of a function is defined by the last return statement it executes. For instance, the function $\lambda:\tau \bullet \acret{\moveop 1}~;~\acret{\moveop 2}$ always returns $2$ by move.
In other words, a function never ``returns early'' and instead ignores non-final return statements.

\subsection{Operational Semantics}

Unlike most variants of $\lambda$-calculi, the $\mathcal{A}$-calculus has to factor the presence of memory into the evaluation of a term, because it has to distinguish between values and references thereupon.
Moreover, as we need to assign capabilities to references, we need a mechanism to identify the latter uniquely.
We achieve this using two levels of indirections.
References are uniquely identified by a reference identifier, which refers to memory location at which semantic values are stored.
Note that reference identifiers cannot be substituted with variable identifiers, because we need to identify anonymous references resulting from function calls or the dereferencing of a record field.
So as to avoid formalizing value representation beyond the strict necessary, with respect to assignment semantics, we neglect data sizes and we assume that a location can hold an arbitrary large value.
We make two additional assumptions on the memory model.
First, we consider memory as an infinite resource, and do not formalize allocation failures.
Second and more importantly, we do not distinguish between heap and stack, and assume memory to be automatically garbage collected.

We define the set of semantic values $\Val$ that describes the data a program may manipulate.
This set includes atomic values, functions and structure instances (a.k.a. records).
While we do treat functions as first-class citizens, we do not let them capture identifiers by closure.
An important consequence of this restriction is that our calculus does not support partial application.
Instead, this can be emulated by the use of functors~\cite[Chapter~22]{Books/Vandevoorde/cpp}.

\begin{definition}[Semantic values]
  Assume an unbounded set of reference identifiers $\Ref$.
  Let $\Atom$ be the set of atomic literals, $\Id$ be the set of identifiers and $\Type$ be the set of types.
  The set of semantic values $\Val$ is given by the minimal set such that:
  \begin{itemize}
    \item
      $\bot \in \Val$ is an undefined value.
    \item
      $a \in \Atom \implies a \in \Val$ is an atomic value.
    \item
      $(X \mapsto \Ref) \in \Val$ is a record, mapping field names from $X \subseteq \Id$ to reference identifiers.
    \item
      $\lambda:\tau~\overline{x} \bullet t \in \Val$ is a function, where $\tau \in \Type$, $\overline{x} \in \Id^{\#}$ and $t \in \AC$.
  \end{itemize}
\end{definition}

\begin{definition}[Type capabilities]
  Let $\Ref$ be the set of reference identifiers.
  The set of type capabilities $\Capa$ is defined as follows:
  \begin{itemize}
    \item $\mathbf{ro} \in \Capa$ denotes the read-only capability.
    \item $\mathbf{rw} \in \Capa$ denotes the read-write capability.
    \item $\mathbf{b}[r] \in \Capa$ denotes the borrowing capability, where $r \in \Ref$ is the reference owning the referred location.
  \end{itemize}
\end{definition}

\begin{definition}[Evaluation context]
  Assume an unbounded set of locations $\Mem$.
  Let $\Val$ denote semantic values,
  an evaluation context $C$ is function with a domain $\{ \nu, \rho, \mu, \kappa \}$, where:
  \begin{itemize}
    \item $C.\nu: \Id \partialto \Ref$ is a table mapping variable names to reference identifiers.
    \item $C.\rho: \Ref \partialto \Mem$ is a table mapping reference identifiers to memory addresses.
    \item $C.\mu: \Mem \partialto \Val$ is a table mapping locations to values.
    \item $C.\kappa: \Ref \partialto \powerset(\Capa)$ is a table mapping reference identifiers to set of type capabilities.
  \end{itemize}
\end{definition}

Let $C$ be a given context, we refer to $C.\nu$ as its \emph{variable table}, to $C.\rho$ as its \emph{reference table}, to $C.\mu$ as its \emph{memory table} and to $C.\kappa$ as its \emph{capability table}.
We say that a context is empty if all its tables are empty (i.e. partial functions with an empty domain).
Furthermore, we use $0 \in \Mem$ to represent the null pointer.

We can finally give the operational semantics of the $\mathcal{A}$-calculus, by the means of big-step inference rules.
All conclusions are of the form $\Gamma, C \vdash t \Downarrow r, C'$, where $C$ and $C'$ are the evaluation contexts before and after evaluating the term $t$, respectively, $r \in \Ref$ is the reference identifier to which the term $t$ evaluates and $\Gamma: \AC \partialto \Type$ is a function that maps terms to their flow-insensitive type.

\subsubsection{Variables Declarations and Dereferencing}
As mentioned earlier, variable declarations do not assign anything to the identifier being declared.
Instead, declarations only consist in associating the variable being declared with a new reference identifier, with no capabilities.
\begin{mathpar}
  \inferrule[E-Let] {
    r_x \not \in\dom(C.\nu) \\
    \Gamma, C[\nu.x \mapsto r_x][\rho.r_x \mapsto 0][\kappa.r_x \mapsto \varnothing] \vdash t \Downarrow r, C'
  }{
    \Gamma, C \vdash \acvar{x:\tau}{t} \Downarrow r, C'
  }
\end{mathpar}
Dereferencing a variable consists in consulting the variable table to find the corresponding reference identifier.
\begin{mathpar}
  \inferrule[E-Var] {
    x \in \dom(C.\nu)
  }{
    \Gamma, C \vdash x \Downarrow C.\nu.x, C
  }
\end{mathpar}

\subsubsection{Assignment Semantics}
The advantage of evaluating all expressions as reference identifiers emerges when formalizing the assignment operators, whose semantics can be expressed purely in terms of reference and memory table updates.

We define a handful of predicates to facilitate the formalization of the type system's constraints.
Given a two reference identifiers $r, s$ and an evaluation context $C$,
a predicate $\ident{contains}(s, r, C)$ indicates whether $s$ denotes a record containing $r$,
a predicate $\ident{readable}(r, C)$ indicates whether $r$ is refers to a readable memory location,
a predicate $\ident{writeable}(r, C)$ indicates whether $r$ is refers to a writeable memory location,
a predicate $\ident{shared}(r, C)$ indicates whether $r$ is shared, and
a predicate $\ident{unique}(r, C)$ indicates whether $r$ is unique.
They are formally defined as follows:
\begin{align*}
  &\ident{contains}(s, r, C) \Leftrightarrow s \in \dom(C.\rho) \land i = C.\mu.(C.\rho.s) \\
    &\quad \land i: X \to L \land \exists x \in X, (i.x = r \lor \ident{contains}(i.x, r)) \\
  &\ident{readable}(r, C)    \Leftrightarrow r \in \dom(C.\kappa) \land C.\kappa.r \cap \{ \mathbf{ro}, \mathbf{rw} \} \not= \varnothing \\
  &\ident{writeable}(r, C)   \Leftrightarrow r \in \dom(C.\kappa) \land C.\kappa.r \cap \{ \mathbf{rw} \} \not= \varnothing \\
    &\quad \land \not\exists s, (\ident{contains}(s, r, C) \land \lnot\ident{writeable}(s, C)) \\
  &\ident{shared}(r, C)      \Leftrightarrow \ident{readable}(r, C) \land \exists s, \mathbf{b}[r] \in C.\kappa.s \\
  &\ident{unique}(r, C)      \Leftrightarrow \ident{readable}(r, C) \land \not\exists s, \mathbf{b}[r] \in C.\kappa.s
\end{align*}
% \begin{align*}
%   \ident{contains}(s, r, C) &\Leftrightarrow s \in \dom(C.\rho) \land i = C.\mu.(C.\rho.s) \\
%                             &\land i: X \to L \land \exists x \in X, i.x = r \\
%   \ident{readable}(r, C)    &\Leftrightarrow r \in \dom(C.\kappa) \land C.\kappa.r \cap \{ \mathbf{ro}, \mathbf{rw} \} \not= \varnothing \\
%   \ident{writeable}(r, C)   &\Leftrightarrow r \in \dom(C.\kappa) \land C.\kappa.r \cap \{ \mathbf{rw} \} \not= \varnothing \\
%                             &\land \not\exists s, \ident{contains}(s, r, C) \land \lnot\ident{writeable}(s, C) \\
%   \ident{shared}(r, C)      &\Leftrightarrow \ident{readable}(r, C) \land \exists s, \mathbf{b}[r] \in C.\kappa.s \\
%   \ident{unique}(r, C)      &\Leftrightarrow \ident{readable}(r, C) \land \not\exists s, \mathbf{b}[r] \in C.\kappa.s
% \end{align*}

Copy assignments consist in taking the value represented by the right operand and store a transitive (a.k.a. deep) copy at the location represented by the left operand.
Therefore, we assume the existence of a function \ident{copy} available at runtime to perform such an operation.
The right operand obviously has to be readable.
As the assignment is mutating, it requires that the left operand be either writeable or unallocated.
\begin{mathpar}
  \mprset{sep=1.5em}
  \inferrule[E-Copy-Mutating] {
    \Gamma, C \vdash t_2 \Downarrow \ident{right}, C_r \\
    \Gamma, C_r \vdash t_1 \Downarrow \ident{left}, C_l \\
    \ident{readable}(\ident{right}, C_l) \\
    \ident{writeable}(\ident{left}, C_l) \\
    l_l = C_l.\rho.\ident{left} \\
    l_r = C_l.\rho.\ident{right} \\
    C' = C_l[\mu.l_l \mapsto \ident{copy}(C_l.\mu.l_r)]
  }{
    \Gamma, C \vdash t_1 \copyop t_2 \Downarrow \ident{left}, C'
  }
\end{mathpar}
If the left operand is unallocated, the assignment must also add a new memory location to memory table's domain.
Furthermore, the capability of the left operand also have to be initialized.
Those will depend on the mutability qualifier of the left operand's type.
If it is declared with the $\acmut$ qualifier, then it will obtain the read-write \emph{and} the read-only capability, otherwise it will obtain the latter only.
We define this principle with a function \ident{ic}, formally described as follows:
\[
  \ident{ic}(\tau) = \begin{cases}
    \{ \mathbf{rw}, \mathbf{ro} \} &\text{if } \acmut \in \tau.q \\
    \{ \mathbf{ro} \} &\text{otherwise}
  \end{cases}
\]
Then we define the evaluation of copy assignments whose left operand is unallocated.
\begin{mathpar}
  \mprset{sep=1.5em}
  \inferrule[E-Copy-Unalloc] {
    \Gamma, C \vdash t_2 \Downarrow \ident{right}, C_r \\
    \Gamma, C_r \vdash t_1 \Downarrow \ident{left}, C_l \\\\
    \ident{readable}(\ident{right}, C_l) \\
    C_l.\rho.\ident{left} = 0 \\
    l_l \not\in \dom(C_l.\mu) \\\\
    l_r = C_l.\rho.\ident{right} \\
    C' = C_l[\mu.l_l \mapsto \ident{copy}(C_l.\mu.l_r)] \\
    C'' = C'[\rho.\ident{left} \mapsto l_l][\kappa.\ident{left} \mapsto \ident{ic}(\Gamma(t_1))]
  }{
    \Gamma, C \vdash t_1 \copyop t_2 \Downarrow \ident{left}, C''
  }
\end{mathpar}
Move assignments are declared similarly, additionally requiring right operands to be unique before destroying their binding and removing their capabilities.
\begin{mathpar}
  \mprset{sep=1.5em}
  \inferrule[E-Move-Mutating] {
    \Gamma, C \vdash t_2 \Downarrow \ident{right}, C_r \\
    \Gamma, C_r \vdash t_1 \Downarrow \ident{left}, C_l \\
    \ident{unique}(\ident{right}, C_l) \\
    \ident{writeable}(\ident{left}, C_l) \\
    l_l = C_l.\rho.\ident{left} \\\\
    l_r = C_l.\rho.\ident{right} \\
    C' = C_l[\mu.l_l \mapsto C_l.\mu.l_r] \\
    C'' = C'[\rho.right \mapsto 0][\kappa.\ident{right} \mapsto \varnothing]
  }{
    \Gamma, C \vdash t_1 \moveop t_2 \Downarrow \ident{left}, C''
  }
\end{mathpar}
\begin{mathpar}
  \mprset{sep=1.5em}
  \inferrule[E-Move-Unalloc] {
    \Gamma, C \vdash t_2 \Downarrow \ident{right}, C_r \\
    \Gamma, C_r \vdash t_1 \Downarrow \ident{left}, C_l \\\\
    \ident{unique}(\ident{right}, C_l) \\
    C_l.\rho.\ident{left} = 0 \\
    l_l \not\in \dom(C_l.\mu) \\\\
    l_r = C_l.\rho.\ident{right} \\
    C' = C_l[\mu.l_l \mapsto C_l.\mu.l_r] \\
    C'' = C'[\rho.\ident{left} \mapsto l_l][\kappa.\ident{left} \mapsto \ident{ic}(\Gamma(t_1))] \\
    C''' = C''[\rho.right \mapsto 0][\kappa.\ident{right} \mapsto \varnothing]
  }{
    \Gamma, C \vdash t_1 \moveop t_2 \Downarrow \ident{left}, C'''
  }
\end{mathpar}
Aliasing assignments consist of an update of the pointer table.
A particular care must be brought to the typestate of the operands, to avoid forming mutating aliases on non-mutating references, and conversely.
To facilitate this task, we define another function $B$ that returns the set of borrowed reference on a particular one, formally described as follows:
\[
  B(r, C) = \{ s \mid s \in \dom(C.\kappa) \land \{ \mathbf{b}[r] \} \in C.\kappa.s \}
\]
Then we define the evaluation of aliasing assignments.
\begin{mathpar}
  \mprset{sep=1.5em}
  \inferrule[E-Alias-Cst] {
    \accst \in \Gamma(t_1).q \\
    \Gamma, C \vdash t_2 \Downarrow \ident{right}, C_r \\
    \Gamma, C_r \vdash t_1 \Downarrow \ident{left}, C_l \\
    \ident{readable}(\ident{right}, C_l) \\
    \lnot\ident{shared}(\ident{left}, C_l) \\
    % \mbox{\small $\ident{shared}(\ident{right}, C_l) \implies \not\exists s, s \not= \ident{right} \land \ident{writeable}(s, C_l)$} \\
    \not\exists s \in B(\ident{right}, C_l), \ident{writeable}(s, C_l) \\
    C' = C_l[\rho.\ident{left} \mapsto C_l.\rho.\ident{right}][\kappa.\ident{left} \mapsto \{ \mathbf{ro}, \mathbf{b}[\ident{right}] \}]
  }{
    \Gamma, C \vdash t_1 \aliasop t_2 \Downarrow \ident{left}, C'
  }
\end{mathpar}
\begin{mathpar}
  \mprset{sep=1.5em}
  \inferrule[E-Alias-Mut] {
    \acmut \in \Gamma(t_1).q \\
    \Gamma, C \vdash t_2 \Downarrow \ident{right}, C_r \\
    \Gamma, C_r \vdash t_1 \Downarrow \ident{left}, C_l \\
    \ident{writeable}(\ident{right}, C_l) \\
    \lnot\ident{shared}(\ident{left}, C_l) \\
    \forall s \in B(\ident{right}, C_l), \ident{writeable}(s, C_l) \\
    C' = C_l[\rho.\ident{left} \mapsto C_l.\rho.\ident{right}][\kappa.\ident{left} \mapsto \{ \mathbf{rw}, \mathbf{b}[\ident{right}] \}]
  }{
    \Gamma, C \vdash t_1 \aliasop t_2 \Downarrow \ident{left}, C'
  }
\end{mathpar}

\subsubsection{Function calls}
We already mentioned that there is a tight connection between assignment semantics and the passing of parameters and return values, as a function is seen as a piece of reusable code that is inlined at each of its call sites.
For example:
\begin{gather*}
  x \moveop (\lambda:\tau y~\bullet \acret{\moveop y})(t~;~y \moveop x) \\
  \equiv \\
  \acvar{y:\tau.dom.y}{t~;~y \moveop x~;~x \moveop y}
\end{gather*}
This resemblance is made explicit in our semantics.
We subdivide function calls' evaluations into three steps:
\begin{description}
  \item[Evaluate the callee]
    This step is described by a dedicated evaluation operator $\Downarrow^{\mathrm{callee}}$
    that retrieves the function value represented by the callee.
    Unlike $\Downarrow$, this operator produces function values (i.e. elements of $\Val$) rather than reference identifiers.
  \item[Evaluate the arguments]
    This step is described by another dedicated evaluation operator $\Downarrow^{\mathrm{args}}$
    that processes argument lists recursively.
  \item[Evaluate the body]
    A function's body is merely a term that should be evaluated in the proper evaluation context.
    Hence this last step is described by $\Downarrow$.
\end{description}
The first step consists in getting the memory location represented by the callee, in order to determine the function's parameters and body.
\begin{mathpar}
  \mprset{sep=1.5em}
  \inferrule[E-Callee] {
    \Gamma, C \vdash t \Downarrow r, C' \\
    \ident{readable}(r, C') \\
    l = C'.\rho.r \\
    C'.\mu.l = \lambda:\tau~\overline{x} \bullet b
  }{
    \Gamma, C \vdash t \Downarrow^{\mathrm{callee}} \lambda:\tau~\overline{x} \bullet b , C'
  }
\end{mathpar}
Recall that we reuse our assignment operators to specify the parameter passing policy.
The alias operator corresponds to the \emph{pass-by-alias} semantics, the mutation operator corresponds to the \emph{pass-by-value} semantics,
and the move operator considers arguments as linear resources.
Consequently, a parameter can be understood as a variable whose declaration and initialization precede the function's body, while an argument describes how its corresponding parameter is initialized.
Therefore, it makes sense to treat each argument as a regular assignment on a freshly declared variable.
\begin{mathpar}
  \mprset{sep=1.5em}
  \inferrule[E-Args-N] {
    \Gamma, C \vdash \acvar{x:\Gamma(x)}{x \lhd t} \Downarrow r_x, C_x \\\\
    \Gamma, C_x \vdash \overline{u} \Downarrow^{\mathrm{args}} C'
  }{
    \Gamma, C \vdash (x \lhd t) + \overline{u} \Downarrow^{\mathrm{args}} C'
  }
  \and
  \inferrule[E-Args-0] {
  }{
    \Gamma, C \vdash \epsilon \Downarrow^{\mathrm{args}} C
  }
\end{mathpar}
As any other expression, a function call should be evaluated as a reference identifier.
Unfortunately, this identifier cannot be known at the call site, because it depends on the return statement that will be executed.
The support for multiple return semantics further complicates the task.
One solution is to take a similar path as that we took for the arguments, and interpret return statements as regular assignments to a ``virtual'' identifier.
Let $\Re \in \Id$ be a reserved identifier denoting a virtual \emph{return identifier}.
Then a return statement is simply an assignment to the return identifier.
In other words, $\acret{\lhd t}$ can be considered equivalent to $\Re \lhd t$.
\begin{mathpar}
  \mprset{sep=1.5em}
  \inferrule[E-Ret] {
    \Gamma, C \vdash \Re \lhd t \Downarrow r, C'
  }{
    \Gamma, C \vdash \acret{\lhd t} \Downarrow r, C'
  }
\end{mathpar}
Sequences of statements are simply executed in order.
Conditional term's evaluation first determines which of the branches should be executed, assuming $\{ \mathtt{true}, \mathtt{false} \} \in \Atom$ are atomic values denoting booleans.
\begin{mathpar}
  \mprset{sep=1.5em}
  \inferrule[E-Seq] {
    \Gamma, C \vdash t_1 \Downarrow r_1, C' \\
    \Gamma, C' \vdash t_2 \Downarrow r_2, C''
  }{
    \Gamma, C \vdash t_1~;~t_2 \Downarrow r_2, C''
  }
\end{mathpar}
\begin{mathpar}
  \mprset{sep=1.5em}
  \inferrule[E-Cond-True] {
    \Gamma, C \vdash t_1 \Downarrow r_1, C' \\
    \ident{readable}(r_1, C') \\
    l = C'.\rho.r_1 \\
    C'.\mu.l = \mathtt{true} \\
    \Gamma, C' \vdash t_2 \Downarrow r_2, C''
  }{
    \Gamma, C \vdash \acif{t_1}{t_2}{t_3} \Downarrow r_2, C''
  }
\end{mathpar}
\begin{mathpar}
  \mprset{sep=1.5em}
  \inferrule[E-Cond-False] {
    \Gamma, C \vdash t_1 \Downarrow r_1, C' \\
    \ident{readable}(r_1, C') \\
    l = C'.\rho.r_1 \\
    C'.\mu.l = \mathtt{false} \\
    \Gamma, C' \vdash t_3 \Downarrow r_3, C''
  }{
    \Gamma, C \vdash \acif{t_1}{t_2}{t_3} \Downarrow r_3, C''
  }
\end{mathpar}
Finally, the rule \TirName{E-Call} assembles all pieces together.
Unfortunately, name shadowing must be taking into account, as function parameters may have the same name as local variables.
A solution is to leverage the $\lambda$-calculus' $\alpha$-conversion, and rename the callee's parameters so that none of them can clash with the identifiers in the current scope.
We describe this process by finding a substitution $\sigma: \Id \partialto \Id$ such that maps shadowing parameters to fresh names.
More formally:
\begin{gather*}
  \sigma = \ident{subst}(\lambda:\tau~\overline{x} \bullet t, C) \\
  \Leftrightarrow \\
  \overline{x} \in \dom(\sigma)^{\#} \land \forall x \in \dom(\sigma), \sigma.x \not\in \dom(C.\nu)
\end{gather*}
Let $\sigma$ be a substitution function and $t \in \AC$ be a term, we write $t[/\sigma]$ the substitution of every identifier $x \in \dom(\sigma)$ by its corresponding image in $\sigma$.
For example $\acvar{x}{x \moveop 2}[/\{ x \mapsto y \}] = \acvar{y}{y \moveop 2}$.
The shadowing of the return identifier also has to be handled, in case it is shadowed by another call inside the function's body.
This is formalized by a function \ident{restore}, defined as follows:
\[
  \ident{restore}(C_a, C') = \begin{cases}
    C'[\nu.\Re \mapsto C_a.\nu.\Re] &\text{if } \Re \in \dom(C_a.\nu) \\
    C' &\text{otherwise}
  \end{cases}
\]
Then we define the evaluation of function calls.
\begin{mathpar}
  \mprset{sep=1.5em}
  \inferrule[E-Call] {
    \Gamma, C \vdash f \Downarrow^{\mathrm{callee}} \lambda:\tau~\overline{x} \bullet t, C_f \\
    \sigma = \ident{subst}(\lambda:\tau~\overline{x} \bullet t, C_f) \\
    (\lambda:\tau'~\overline{x}' \bullet u') = (\lambda:\tau~\overline{x} \bullet t)[/\sigma] \\
    \Gamma' = \Gamma[\sigma.x \mapsto \Gamma.x \mid x \in \dom(\sigma)] \\
    \Gamma', C_f \vdash \overline{u}[/\sigma] \Downarrow^{\mathrm{args}} C_a \\
    \tau_\Re = \tau'.codom \\
    \Gamma'[\Re \mapsto \tau_\Re], C_a \vdash (\acvar{\Re:\tau_\Re}{t'}) \Downarrow r, C' \\
    C'' = \ident{restore}(C_a, C')
  }{
    \Gamma, C \vdash f(\overline{u}) \Downarrow r, C''
  }
\end{mathpar}

\subsubsection{Values}

Literal values are allocated to a new memory location, to which a reference identifier is created, associated with the capabilities corresponding to the term's type.
\begin{mathpar}
  \mprset{sep=1.5em}
  \inferrule[E-Atom] {
    a \in \Atom \\
    r \not\in \dom(C.\nu) \\
    l \not\in \dom(C.\mu) \\
    C' = C[\rho.r \mapsto l][\mu.l \mapsto a][\kappa.r \mapsto \ident{ic}(\Gamma(a))]
  }{
    \Gamma, C \vdash a \Downarrow r, C'
  }
  \\
  \inferrule[E-Fun] {
    r \not\in \dom(C.\nu) \\
    l \not\in \dom(C.\mu) \\
    C' = C[\rho.r \mapsto l][\mu.l \mapsto \lambda:\tau~\overline{x} \bullet t][\kappa.r \mapsto \ident{ic}(\tau)]
  }{
    \Gamma, C \vdash \lambda:\tau~\overline{x} \bullet t \Downarrow r, C'
  }
\end{mathpar}
Records are instantiated by creating a reference identifier for each of their fields, stored as a function mapping the corresponding field names to them.
This essentially creates a separate variable table, specific to the record's fields.
Therefore, dereferencing a field is similar to how variables are dereferenced, only differing in the table that is consulted.
\begin{mathpar}
  \mprset{sep=1.5em}
  \inferrule[E-New] {
    \forall x \in \dom(\tau), r_x \not\in \dom(C.\nu) \\
    i = \tuple{x \mapsto r_x \mid x \in \dom(\tau)} \\
    r \not\in \dom(C.\nu) \\
    l \not\in \dom(C.\mu) \\
    C' = C[\rho.r \mapsto l][\mu.l \mapsto i][\kappa.r \mapsto \ident{ic}(\tau)] \\
    C'' = C'[\rho.r_x \mapsto 0 \mid x \in \dom(\tau)][\kappa.t_x \mapsto \varnothing \mid x \in \dom(\tau)]
  }{
    \Gamma, C \vdash \acnew{\tau} \Downarrow r, C''
  }
\end{mathpar}
\begin{mathpar}
  \mprset{sep=1.5em}
  \inferrule[E-Field] {
    \Gamma, C \vdash t \Downarrow r, C' \\
    \ident{readable}(r, C') \\
    l = C'.\rho.r \\
    x \in \dom(C'.\mu.l)
  }{
    \Gamma, C \vdash t.x \Downarrow C'.\mu.l.x, C'
  }
\end{mathpar}
\section{Related Work}

While many contemporary programming languages offer a tight control over the assignment semantics (e.g. Rust or C/C++), very few propose different assignment operators rather than overloading a single one with different semantics.
The R language proposes two operators, but those are related to variable visibility rather than assignment semantics~\cite[Chapter~2]{Paradis:2002:R}.
The Go language also features multiple operators, for similar purposes~\cite[Chapter~2]{DBLP:journals/software/Meyerson14}.
Highly customizable languages (e.g. Python, C++ or Swift) can let augmented assignment operators such as ``\lstinline|+=|'' or ``\lstinline|-=|'' be redefined to suit a custom assignment semantics.

Early proposals to formalize assignments semantics include $\lambda_{var}$~\cite{DBLP:conf/popl/OderskyRH93}, an extension of the classic $\lambda$-calculus that features assignable variables.
These are introduced into an expression by an additional construct $\acvar{x}{e}$, which delimits their lexical scope.
% They only differ from regular variables in that they can appear on the left operand of an assignment.
% Sequences of instructions are handled with monads.
A subsequent line of research focus on equipping $\lambda_{var}$ with type systems (e.g~.\cite{TR:Hongseok:1997,DBLP:conf/tacs/ChenO94}) to guarantee freedom from side effects for pure applicative terms.
Another approach from the same era formalizes the notion of memory location, in order to support a call-by-value approach~\cite{DBLP:journals/tcs/FelleisenF89}.
Similar to $\lambda_{var}$ the authors propose assignable variables, but those are interpreted as memory locations in terms' reductions.
This semantics is revisited in ClassicJava~\cite{DBLP:conf/popl/FlattKF98} to formalize a subset of Java.
The resounding success of object-orientation in the last couple of decades has pushed into the direction of object-oriented calculi.
Among the most popular is Featherweight Java~\cite{DBLP:journals/toplas/IgarashiPW01}, that aims to provide a minimal calculus for studying the consequences of extensions to the Java language.
Evidently, support for assignments is a natural extension for Featherweight, and is proposed in a variety of models (e.g.~\cite{DBLP:conf/ecoop/MackayMPGC12,TR:Bierman2003MJ}).
More recently, \cite{DBLP:journals/entcs/CapriccioliSZ16} introduces a calculus whose semantics is not defined with an auxiliary memory structure to represent assignable variables' bindings.
Instead, the authors proposes to stay at a higher abstraction level and reduce assignable variables to the value they represent.
The advantage is that aliasing is conveyed at a syntactic level, whereas approaches based on an external store require aliasing information to be derived from variable mappings.
The model is further extended in~\cite{DBLP:journals/corr/abs-1904-10107} to integrate aliasing control mechanisms.
In comparison, our approach aims to model memory more precisely, at the cost of a more complex operational semantics.

Our type system heavily borrows from the domain of typestate analysis~\cite{DBLP:journals/tse/StromY86} and type capabilities~\cite{DBLP:conf/ecoop/BoylandNR01}.
More specifically, our treatment of non-mutating references is reminiscent to~\cite{DBLP:conf/popl/NadenBAB12}.
Our work is also related to Ownership Types~\cite{DBLP:series/lncs/ClarkeOSW13}.
While Ownership Types were originally designed to control the exposure of an object's internal representation, a plethora of variants have been to proposed to address different issues, including mutability~\cite{DBLP:journals/toplas/DietlDM11}.
In particular, Anzen's restrictions on non-mutating aliases to mutating fields is reminiscent to the \emph{owners-as-accessors} strategy~\cite{DBLP:conf/icse/PotaninDN13}, which proscribes access to a reference unless its owner appears on the call stack (similar to our example involving inversion of control).
The main difference between these approaches and ours is that Anzen performs typestate checking at runtime rather than statically.
While this induces an overhead on execution, it allows for a more precise tracking of uniqueness and immutability by avoiding conservative hypotheses.

\section{Concluding Remarks}

We have presented Anzen, a general purpose language that aims to make assignments easier to understand and manipulate.
Although it supports multiple paradigms, Anzen leans towards object orientation and imperative programming.
The most distinguishing feature of the language is its three different assignment operators, that together can reproduce all common imperative assignment strategies explicitly.
One describes aliasing, another describes cloning and the last relates to assignments that preserve uniqueness.
% To the best of our knowledge, Anzen is the first attempt at using different operators to control the way memory is bound to references.
The language is also equipped with a type system to control uniqueness and immutability dynamically, based on type capabilities.
Constraints are specified using type annotations on variable declarations and function signatures.

We have introduced Anzen's features informally, through various simple examples, and illustrated their use to solve a non-trivial problem that caused a security breach in a past implementation of Java.
Then, we have formalized Anzen's semantics and type system by the means of a minimal formal language called the $\mathcal{A}$-calculus.
An implementation of Anzen's interpreter is distributed as an open-source software and available at~\url{https://github.com/anzen-lang/anzen}.

While we have presented our approach through the lens of a particular language, Anzen's operators and type systems can be adapted to other languages, as we did in~\cite{DBLP:conf/sle/RacordonB18} for a dialect of JavaScript.

\bibliography{references}

\end{document}